# Resonate and Fire Neuron with Fixed Magnetic Skyrmions

Md. Ali Azam[1], Dhritiman Bhattacharya[1], Damien Querlioz[2], Jayasimha Atulasimha[1,3,*]

1. Department of Mechanical and Nuclear Engineering, Virginia Commonwealth University.
2. Centre de Nanosciences et de Nanotechnologies, CNRS, Univ Paris-Sud, Université Paris-Saclay, 91405 Orsay France.
3. Department of Electrical and Computer Engineering, Virginia Commonwealth University.

*jatulasimha@vcu.edu

**Abstract:** In the brain, the membrane potential of many neurons oscillates in a subthreshold damped fashion and fire when excited by an input frequency that nearly equals their eigen frequency. In this work, we investigate theoretically the artificial implementation of such "resonate-and-fire" neurons by utilizing the magnetization dynamics of a fixed magnetic skyrmion in the free layer of a magnetic tunnel junction (MTJ). To realize firing of this nanomagnetic implementation of an artificial neuron, we propose to employ voltage control of magnetic anisotropy or voltage generated strain as an input (spike or sinusoidal) signal, which modulates the perpendicular magnetic anisotropy (PMA). This results in continual expansion and shrinking (i.e. breathing) of a skyrmion core that mimics the subthreshold oscillation. Any subsequent input pulse having an interval close to the breathing period or a sinusoidal input close to the eigen frequency drives the magnetization dynamics of the fixed skyrmion in a resonant manner. The time varying electrical resistance of the MTJ layer due to this resonant oscillation of the skyrmion core is used to drive a Complementary Metal Oxide Semiconductor (CMOS) buffer circuit, which produces spike outputs. By rigorous micromagnetic simulation, we investigate the interspike timing dependence and response to different excitatory and inhibitory incoming input pulses. Finally, we show that such resonate and fire neurons have potential application in coupled nanomagnetic oscillator based associative memory arrays.

## I. Introduction

Following the pioneering vison of Carver Mead [1], neuromorphic computing has garnered considerable interest in recent times due to its potential advantage in dealing with computational problems with ill conditioned input data, adaptive nature of these systems to mitigate the effect of component failure and efficiency compared to fully Boolean logic based computation [2–5]. Due to the complex and mixed analog-digital nature of the brain, a major hurdle towards developing neuromorphic computation platforms has been finding materials and devices to mimic brain like behavior and developing architectures based on such systems. Current hardware artificial neural networks are mostly implemented with purely CMOS circuits and require large number of components to ensure robustness [3,6]. This poses a major challenge for the scaling and energy efficiency of neuromorphic computation.

Nanomagnetic devices are one of the promising alternatives to implement neuromorphic computing and other non-von-Neumann like architectures due to their low energy consumption, nonlinear dynamics, and non-volatility. Many nanomagnetic devices that can potentially form the building blocks of neuromorphic computing: artificial neurons and synapses, have been proposed [4,7–14]. Among artificial neurons, most emulate the behavior of (leaky) integrate and fire type neurons [3,10]. In an integrate and fire type neuron, the membrane potential increases in response to an input spike and fires if it reaches a certain threshold [15]. Therefore, the firing frequency depends only on the strength of the stimulus. However, in the brain, many neurons also feature damped or sustained subthreshold oscillation [16–19] of membrane potential. Such neurons therefore show sensitivity towards the timing of stimulus. Consequently, a strong stimulus may not produce a spiking output if the incoming stimulus is not in phase with the oscillation of membrane potential, thus providing an inhibitory function. It also leads to many interesting phenomena such as fluctuation of spiking probability and selective communication [20]. Such "resonate-and-fire" neurons could also be useful in different neural networks where computation involves synchronized oscillation of several spin torque nano-oscillators (SNTOs) for pattern recognition [21]. Such networks come in different versions: for example, in ref. [21] patterns are encoded by frequency-shift keying (FSK) whereas in most other work [22–24] patterns are encoded with phase shift keying (PSK), but all of them could benefit from circuits able to detect synchrony through resonance.

In this work, we investigate the implementation of an artificial resonate-and-fire neuron by utilizing the magnetization dynamics of a fixed magnetic skyrmion in the free layer of a magnetic tunnel junction. Magnetic skyrmions (Fig. 1 (a)) are topologically protected spiral spin textures [25,26], which can be translated by applying small current [27,28] or reversed (in patterned dots) using a small voltage that controls the magnetic anisotropy [29,30]. Until now, this behavior has been leveraged to propose logic and memory devices based on magnetic skyrmions [31–35].

While nanomagnetic device based integrate and fire type neurons have been studied extensively, the resonate and fire type neuron proposed using nanomagnetic devices in this work, is unique to this paper and would be essential to compare the synchronization in arrays of nanomagnetic oscillators as described above for applications such as pattern recognition. While various schemes mimicking neuron and synapse activities have also been proposed utilizing current induced motion of skyrmions [36–39], neuromorphic devices based on moving skyrmions could have a large foot print and are dissipative as they use current to move the skyrmions. We previously proposed nanomagnetic memory devices utilizing voltage control of fixed magnetic skyrmions in the free layer of a MTJ structure [29,40,41] that can alleviate these issues. In this paper, we use such a voltage control of a fixed skyrmion based scheme to achieve the functionality of resonate and fire neurons.

The next section describes two "resonate and fire" neuron devices based on the novel mechanism of resonant oscillations of a skyrmion core due to voltage control of anisotropy: direct voltage control of magnetic anisotropy (VCMA) and strain mediated voltage control of anisotropy in magnetostrictive materials. This is followed by a section explaining the modeling of voltage induced magnetization dynamics, followed by a discussion of skyrmion core oscillation, resonant behavior and application of the "resonate and fire" functionality for detection of phase and frequency synchronization.

## II. Device:

Our proposed device is a MTJ structure in which the circular free layer hosts a fixed skyrmion. We propose two different devices where application of a voltage modulates the perpendicular anisotropy of the free layer through two different physical mechanisms. The anisotropy can be modulated via voltage control of magnetic anisotropy [42–44] in the device shown in Figure 1(b) and voltage generated strain [45,46] in the device shown in Figure 1(c). Modulation of perpendicular anisotropy in the system induces breathing of skyrmions. In other words, the skyrmion core increases and decreases in size. This mimics the subthreshold damped oscillations of resonate and fire neurons. The electrical resistance of the MTJ layer ($R_2$) depends on the magnetization orientation of the free layer (i.e. the size of the skyrmion core) relative to that of the fixed layer. For the sake of explanation, let us assume that the orientation of the magnetization of the fixed layer is antiparallel with respect to the one of the skyrmion core. As the skyrmion core expands during the breathing, more spins in the free layer will be antiparallel with respect to the fixed layer spins. Therefore, the resistance of the MTJ structure will increase. We thus propose to use a voltage divider consisting of a fixed resistor and the voltage controlled MTJ resistance to drive a CMOS buffer from OFF to ON state as shown in Fig 1.

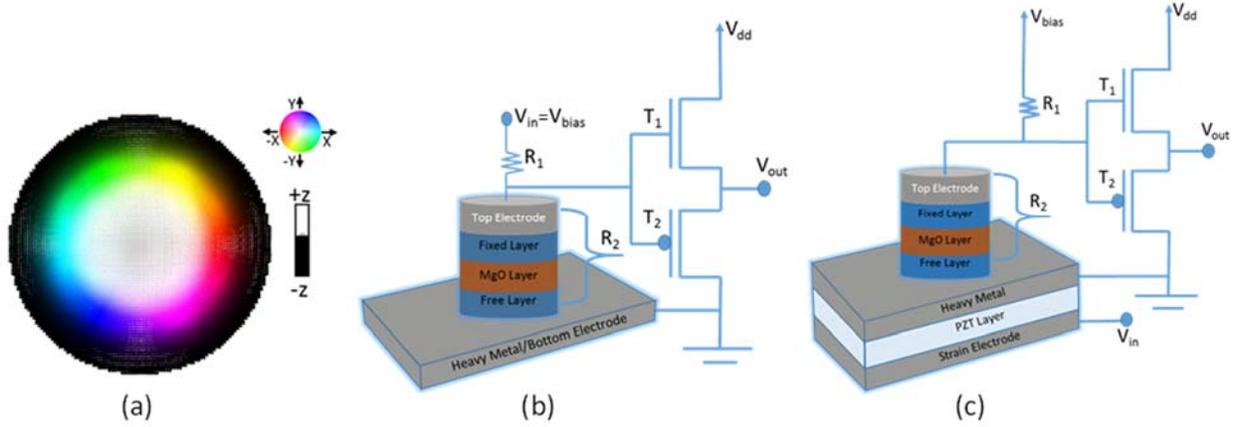

Figure 1. (a) A core-up skyrmion, color code on the right shows the direction of the spin, (b) Proposed device structure operated with voltage control of magnetic anisotropy (VCMA) (c) MTJ structure stacked on PZT layer for strain control of magnetic anisotropy. Note that, CMOS buffer is driven by the MTJ resistance. Therefore, fixed layer magnetization needs to be opposite to that of skyrmion core.

If the resistance of the MTJ stack increases during skyrmion expansion, potential drop across the MTJ resistance ($R_2$) will be higher. We can choose the ratio between $R_1$ and $R_2$ such that, at a given threshold this increase causes the transistor T1 to be turned on and generate a firing pulse. (NOTE: A similar behavior can be achieved by choosing the fixed layer magnetization orientation to be parallel with respect to the magnetization orientation of the skyrmion core and driving the CMOS buffer by the potential drop across resistance $R_1$). In short, if the skyrmion core size increases beyond a threshold, potential drop across the MTJ stack will produce an input voltage to the CMOS circuit that will exceed the threshold voltage, causing the buffer to "fire", i.e. produce a high output. In ref. [47] resistance-area product of MTJ was found to be in the range of 225-650 $\Omega.\mu m^2$ and typical tunnel magneto-resistance ratio between parallel and antiparallel configuration is 100% [47]. However, in this case, magnetization is oscillating between a skyrmionic state and a quasi-antiparallel state. Typical CMOS buffer have gating voltage in the range of 1 V. Hence, we can design a bias voltage ($V_{bias}$) and appropriate ratio for $R_1$ to $R_2$. It would be preferable to maximize $R_1$ and $R_2$ (to minimize standby power dissipation due to $V_{bias}$) while ensuring reasonable RC time constant for resonant operation of the device. In this work, for the sake of simplicity, we do not model the magnetoresistance change due to skyrmion breathing and the circuit dynamics of the CMOS buffer. In other words, in Fig 1 (corresponding to both device implementations for the resonate and fire neuron), we only model the magnetization dynamics of the skyrmions and set a threshold value of average magnetization along the z-axis ($m_{z\_threshold}$=0.8, magnetization is almost antiparallel to the free layer). For $m_{z\_free}$> $m_{z\_threshold}$, we consider the CMOS buffer to be in the 'ON' or "high" state and "OFF" or "low" otherwise. This naturally gives rise to a firing output.

## III. Methods

Micromagnetic simulation software MuMax3 [48] was used to perform the simulations where the magnetization dynamics is found by solving the Landau-Lifshitz-Gilbert (LLG) equation,

$$\frac{\partial \vec{m}}{\partial t} = \vec{\tau} = \left(\frac{-\gamma}{1+\alpha^2}\right)\left(\vec{m} \times \vec{H}_{eff} + \alpha\left(\vec{m} \times \left(\vec{m} \times \vec{H}_{eff}\right)\right)\right) \quad (1)$$

where $\vec{m}$ is the reduced magnetization ($\vec{M}/M_s$), $M_s$ is the saturation magnetization, $\gamma$ is the gyromagnetic ratio and $\alpha$ is the Gilbert damping coefficient. The effective magnetic field $\vec{H}_{eff}$ is given by,

$$\vec{H}_{eff} = \vec{H}_{demag} + \vec{H}_{exchange} + \vec{H}_{anis} + \vec{H}_{thermal} \tag{2}$$

Here, $\vec{H}_{demag}$ is the effective field due to demagnetization energy, $\vec{H}_{exchange}$ is the effective field due to Heisenberg exchange coupling and Dzyaloshinskii-Moriya Interaction (DMI). The DMI contribution to $\vec{H}_{exchange}$ is given by [48]:

$$\vec{H}_{DM} = \frac{2D}{\mu_0 M_s}\left[\frac{\partial m_z}{\partial x}, \frac{\partial m_z}{\partial y}, -\frac{\partial m_x}{\partial x} - \frac{\partial m_y}{\partial y}\right] \tag{3}$$

where $m_z$ is the z-component of magnetization and D is the effective DMI constant.

The effective field due to the perpendicular anisotropy, $\vec{H}_{anis}$, is expressed as [48],

$$\vec{H}_{anis} = \frac{2K_{u1}}{\mu_0 M_s}(\vec{u}.\vec{m})\vec{u} + \frac{4K_{u2}}{\mu_0 M_s}(\vec{u}.\vec{m})^3\vec{u} \tag{4}$$

where $K_{u1}$ and $K_{u2}$ are first and second order uniaxial anisotropy constants and $\vec{u}$ is the unit vector in the direction of the anisotropy (i.e. perpendicular anisotropy in this case). VCMA/strain effectively modulates the anisotropy energy density. The resultant change in uniaxial anisotropy due to VCMA/strain is incorporated by modulating $K_{u1}$ while keeping $K_{u2} = 0$. For VCMA, this change is given by $\Delta k_{u1} = \Delta \text{PMA} = aE$. Here, $a$ and E are respectively the coefficient of electric field control of magnetic anisotropy and the applied electric field. On the other hand, for strain, this is given by $\Delta k_{u1} = \Delta \text{PMA} = \frac{3}{2}\lambda\sigma$, where $\lambda$ and $\sigma$ are respectively the magnetostriction coefficient and the applied stress.

| Parameters | Value |
|---|---|
| Saturation Magnetization ($M_{sat}$) | $1\times 10^6$ A/m |
| Exchange Constant ($A_{ex}$) | $2\times 10^{-11}$ J/m |
| Perpendicular Anisotropy Constant ($K_{u1}$) | $6\times 10^5$ J/m$^3$ |
| Gilbert Damping ($\alpha$) | 0.03 |
| DMI Constant (D) | 3 mJ/m$^2$ |

In order to reduce the effect of VCMA/strain on the fixed layer, the thickness of the fixed layer can be made lower compared to that of the free layer. This lower thickness ensures a high perpendicular magnetic anisotropy. Materials with low VCMA/magnetostriction co-efficient can be chosen for the fixed layer so that effect of voltage application is minimal. Additionally, one can use a synthetic antiferromagnetic [49] layer to increase magnetic stability of the fixed layer. These ensure the magnetization of the fixed magnetic layer does not rotate much due to VCMA or due to strain (if any) transferred to it. Therefore, we only simulate the magnetization dynamics of the free layer. The synthetic antiferromagnetic layer also offsets the dipolar interaction between the fixed and the free layer. Hence, we ignore anti-symmetric modification effects due to dipolar effects in our model. Exchange interaction and DMI can be modulated when an electric field is applied. However, these effects are minimal [41] and will only result in a small change in the breathing frequency and will not change the key results of our work significantly.

## IV. Results

## A. Damped Oscillatory Behavior of Skyrmions

We simulated the magnetization dynamics in a 100 nm diameter nanodisk with thickness of 1 nm. Our geometry was discretized into $1\times 1\times 1$ $nm^3$ cells. Using the parameter values in table I, the ground magnetization state was found to be a skyrmion. A triangular input spike of $\Delta PMA=1\times10^5$ J/m³ was applied with 50 ps rise and 50 ps fall time (NOTE: We use fast rise and fall time in the triangular pulse to simulate response to a near ideal pulse whereas sinusoidal inputs are used later for more realistic device simulations). Furthermore, we are mostly interested in frequency or phase synchronization of sinusoidal waveforms, but nevertheless choose triangular spikes initially due to similarity to actual spike like stimulus available in real neurons (though time scales for biological and artificial skyrmions resonate and fire neurons are vastly different). The momentary change in anisotropy causes the core of the skyrmions to expand and oscillate about the equilibrium state.

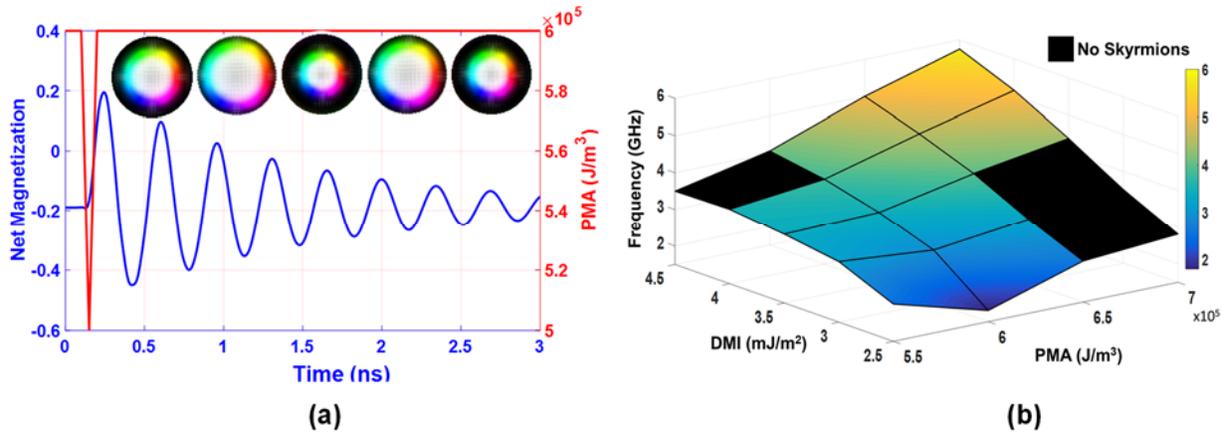

Figure 2. (a) Damped oscillation of a fixed skyrmion's core due to stimulation with a single pulse [Red color line: Input spike, Blue color line: Output average magnetization along the perpendicular direction (z-axis) (b) Modulation of breathing frequency by varying the interfacial parameters.

The oscillatory behavior can be seen from the net magnetization curve in Figure 2 (a). This imitates the subthreshold neuron oscillation of a resonant neuron. From this magnetization dynamics, the breathing frequency of the skyrmion can be determined. This is analogous to the eigen frequency (i.e. damped oscillation frequency) of the neuron. This information is important as an input spike train or sinusoid should have a frequency that nearly equals the eigen frequency to cause a neuron to resonate and spike. This breathing frequency strongly depends on the parameters of the system. Here, we determine the breathing frequency as a function of interfacial parameters PMA and DMI (Fig. 2 (b)). This frequency can be easily tuned in the range of 1.8 GHz to 5.75 GHz. In addition to interfacial parameters, one can use a DC bias voltage to change the PMA which will subsequently tune the frequency about which the skyrmion oscillates but this is not discussed here as it is beyond the scope of this paper.

## B. Resonant behavior of Skyrmions

The skyrmion breathing frequency estimated in the last sub-section is now utilized to drive the skyrmion into resonance and show the resonate and fire behavior is very sensitive to this excitation frequency. We again start with triangular input pulses for reasons mentioned in the prior sub-section. Triangular pulses of $\Delta PMA=1.5\times10^5$ J/m³ of time interval in a range of 0.35-0.50 ns was applied to the system. At the PMA chosen (in the absence of voltage applied) the core had a breathing frequency of approximately 2.86 GHz (T=0.35 ns). Skyrmion breathing of significant amplitude was observed when two input spikes were separated by an interval that falls in the range 0.43 ns to 0.46 ns. Breathing with diminishing amplitude was observed in other cases. The example in Fig. 3(a) shows cases for 3 different time intervals between two successive input spikes: 0.35 ns, 0.45 ns and 0.50 ns.

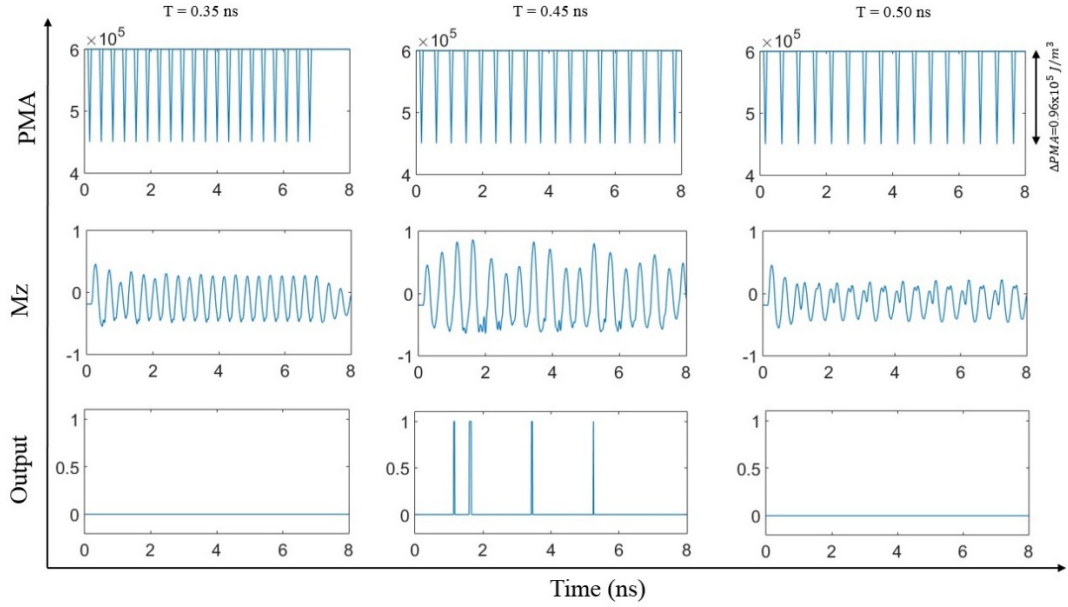

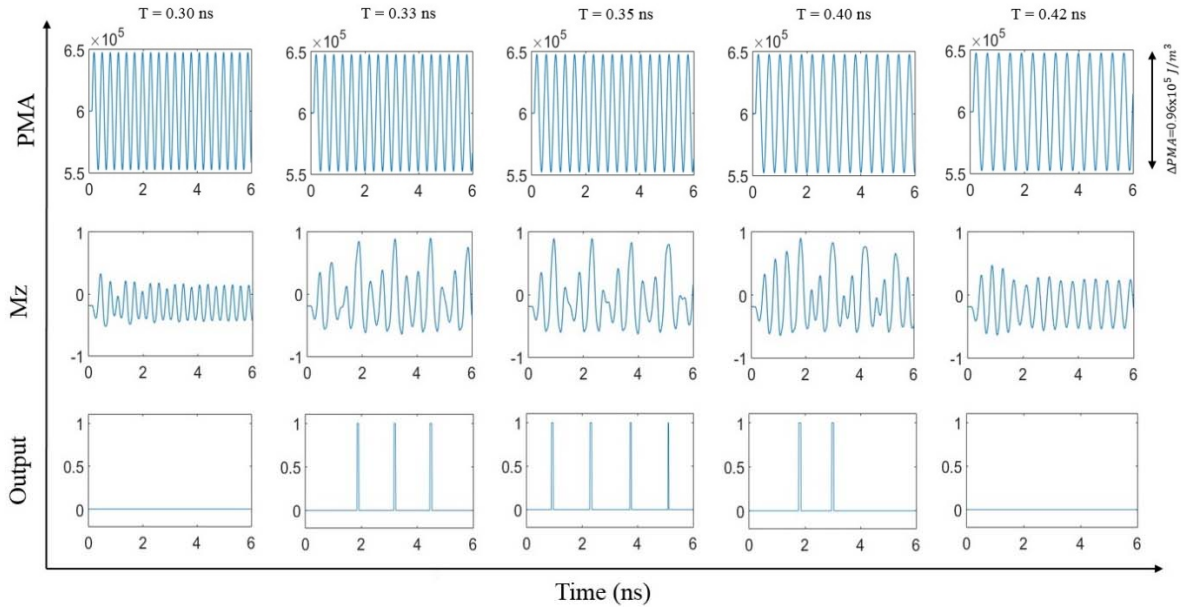

Figure 3. Resonant behavior: (a) Spike input (b) sinusoidal input

Considering M_z_threshold=0.8, a spiking output can be found for time interval of 0.45 ns, while no output spikes are found when the time interval was 0.35 ns and 0.50ns. Other than the dependency on time interval the skyrmion core resonance is significantly sensitive to the amplitude of the input impulse. Lowering ΔPMA to $1.4\times10^5$ J/m³ failed to produce any output as expected due to sub-threshold oscillation. However, increasing the ΔPMA to $1.75\times10^5$ J/m³ lowered (instead of increasing) the firing rate from 4 for first 10ns to just 2. This is because we consider M_z_threshold=0.8, which occurs when ~80% of the spins point upwards. As the core size is very large, peripheral (boundary) effects strongly influence the breathing dynamics which makes the behavior strongly nonlinear. Due to this, the correlation between the change of input magnitude and the spiking behavior is hard to predict at these limits. Hence, for triangular input with T=0.45ns, ΔPMA of $1.5 \times 10^5$ J/m$^3$ resulted in the best firing behavior.

While triangular spikes were used to illustrate the spiking behavior, a sinusoidal input pulse is more useful for many practical applications. Appropriate frequency sinusoidal inputs can also result in firing due to the same principle, i.e. a sinusoid of given amplitude, whose frequency is resonant with the eigen frequency produces the strongest spiking behavior. Sinusoids of different frequencies with peak to peak ΔPMA=$0.96 \times 10^5$ J/m$^3$ were used as inputs. Strongest firing (4 spikes over 6 ns) was found around 2.86 GHz (time period of 0.35 ns) input frequency. Higher frequency (3 GHz or time period of 0.33 ns) and lower frequency (2.5 GHz or time period of 0.4 ns) resulted in weaker spiking behavior (less than 4 spikes over the same 6 ns). Further deviation in frequency from resonance: 3.3 GHz (time period of 0.3 ns) and 2.38 GHz (time period of 0.42 ns) resulted in no spiking behavior at all.

We note that the eigen frequency (for single triangular pulse) and resonant frequency for triangular and sinusoidal inputs all appear to be different. This is because the ΔPMA produced by the input voltage leads to a variation in the net PMA experienced by the breathing skyrmion, which in turn changes its frequency.

The change in perpendicular anisotropy is given by ΔPMA = $a$E. The VCMA co-efficient was found to be as large as 290 fJ/Vm experimentally[50] and 1800 fJ/Vm theoretically[51]. Using a= 100 fJ/Vm, peak to peak ΔPMA=$0.96 \times 10^5$ J/m$^3$ can be achieved using a peak voltage of 0.48 V, considering the MgO layer to be of 1 nm thickness. Energy dissipation per cycle will be $\frac{1}{2}CV^2$ considering energy dissipation is dominated by the energy required to charge the capacitive MgO layer. The capacitance of the MgO layer C= 0.487 fF for relative permittivity of 7 [52]). This translates into an energy dissipation of 56 aJ.

On the other hand, change in PMA achieved via strain is given by $\Delta PMA = \frac{3}{2}\lambda\sigma$. Magnetostrictive coefficient $\lambda$ is 37 ppm for CoFeB [53]. Stress cycles with magnitude ~1 GPa will be needed to drive this system to resonance, which is not practical. Materials with higher magnetostrictive coefficients exist. For example, FeGa has a coefficient of 300 ppm [54] while $\lambda$ can be as high as 1000 ppm for Terfenol-D [55,56]. The requirement of stress will be correspondingly lower (respectively ~167 MPa and ~50MPa) assuming that such material systems with such highly magentostrictive materials also exhibit DMI (has not been studied so far). To generate 50 MPa stress, required voltage is 83.375 mV. We again consider energy dissipation is dominated by the energy required to charge the capacitive piezoelectric layer. The relative permittivity of the piezoelectric layer is taken to be 1000. Considering 100 nm thick PZT layer, capacitance C=0.695 fF. This gives rise to an energy dissipation of mere 2.4 aJ.

We note that, energy dissipation ~ femto-Joules (fJ) in the resistive elements (due to V$_{bias}$) will dominate energy dissipated in the scaled MTJ (~10-100 aJ) as well as the CMOS buffer (each CMOS device typically require ~100 aJ per switching event [57]). Thus, the total energy requirement will be ~ femto-Joule/spiking event.

These values are highly attractive in comparison to a purely CMOS implementation of the resonate and fire neuron. In reference [58], CMOS implementation of a resonate-and-fire neuron involves capacitors approaching pico-Farad, leading to an energy consumption per firing event in the range of pico-Joules, an area of many micrometer square and resonant frequency of a few 10s of Hz. In fact, the proposed hybrid skyrmion-MTJ and CMOS buffer implementation of the resonate and fire neuron, is capable of resonant frequencies ~few GHz and is potentially 3 orders of magnitude more energy efficient/spiking event and potentially has 2 orders of magnitude higher density than that the all CMOS implementation [58] as shown in Table 1.

Table 1. Performance comparison of proposed hybrid nanomagnet-CMOS vs. all CMOS resonate and fire neuron [58].

| Performance metric | Hybrid fixed skyrmion-MTJ and CMOS buffer | All CMOS [58] |
|---|---|---|
| Energy dissipation/spiking event | ~ femto-Joule | ~pico-Joule |
| Density (area per device) | ~0.01 micron$^2$ | ~micron$^2$ |
| Resonance frequency | ~ GHz | ~10s Hz (can be designed to be much faster) |

## C. Frequency and Phase synchronization detection of STNO oscillators

Frequency and phase synchronization detection of coupled spin torque nano-oscillators (STNO) is an important component is neuromorphic computing schemes that implement associative memory[22–24]. In this section, we show that our proposed device can be used to detect phase and frequency synchronization of STNOs and, in general, any coupled oscillators. In this subsection, we consider the outputs of two STNOs (or two coupled oscillators) in general have been added together as:

$$V = V\sin(2\pi f_1 + \phi_1) + V\sin(2\pi f_2 + \phi_2)$$

Here $f_1$ ($f_2$) and $\Phi_1$ ($\Phi_2$) are respectively the frequency and phase of the first and second oscillator output.

**Case I: Phase differs, frequency synchronized**

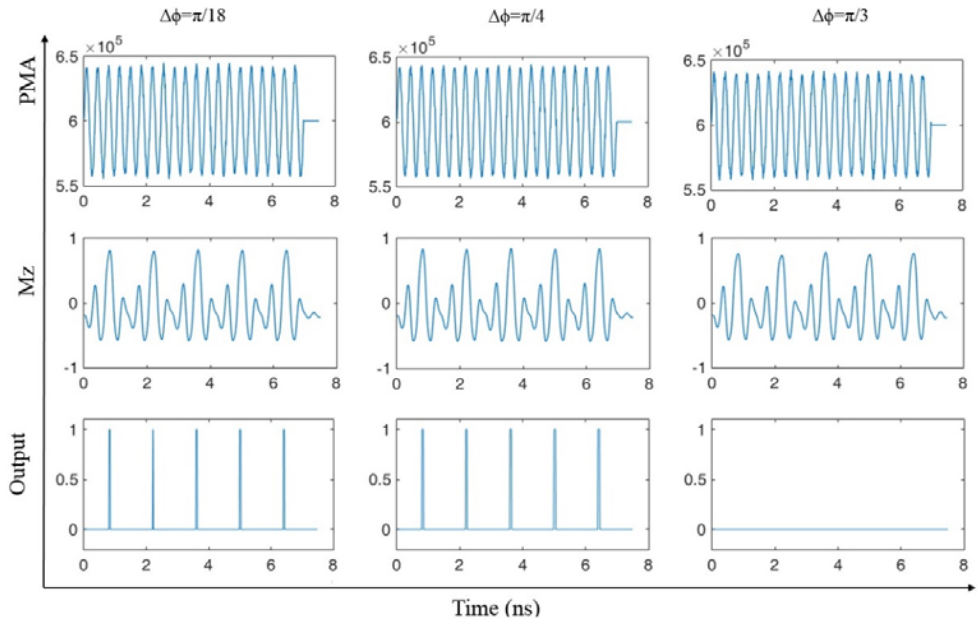

Figure 4. Phase detection: As the phase difference increases the amplitude of the input decreases thus making it harder for the magnetization to reach the threshold limit for firing

Here the two signals have no difference in frequency but have a phase difference (Δ∅) of pi/18 (10°), pi/4(45°), pi/3 (60°). We also include a random phase noise as follows:

$$\emptyset_1 = \emptyset_{1_0} + \emptyset_{random,1}$$

$$\emptyset_2 = \emptyset_{2_0} + \emptyset_{random,2}$$

$$\Delta\emptyset = \emptyset_1 - \emptyset_2$$

The random phase noise added here is white noise. In figure 4 we show that output spikes several times when the phase difference is below a certain limit (e.g. 45° and below) and when the phase difference is larger (60°) the output fails to spike.

**Case II: Frequency differs, (we assume that at t=0, Δ∅ =0)**

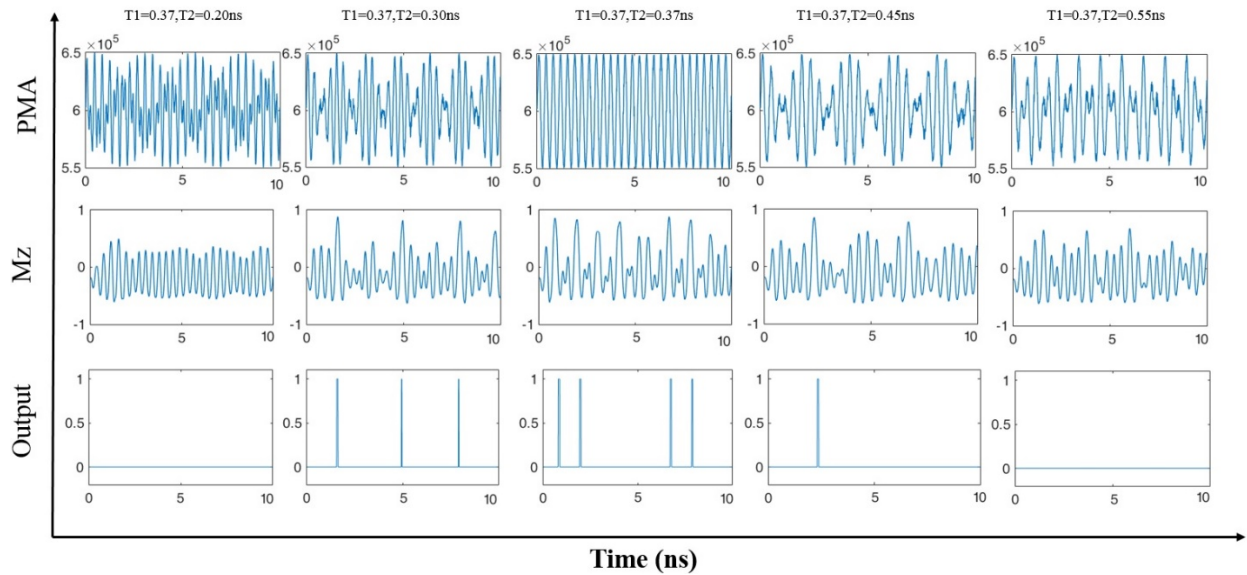

Figure 5. Frequency synchronization detection

Here the two signals have different frequency but have no phase difference at t=0. Both signals are subjected to phase noise and the spiking output is analyzed over 10 ns. A signal with a frequency of 2.7 GHz (T=0.37ns) is chosen as the base signal. This frequency is slightly lower than the actual resonance frequency (2.86GHz) and intentionally chosen so to demonstrate robustness of the frequency synchronization detection to frequencies that are slightly off resonance. Successive signals added to it have frequencies of 5 GHz, 3.33 GHz, 2.7 GHz, 2.22 GHz and 1.82 GHz. When both frequencies are equal (2.7 GHz) 4 spikes are produced in 10 ns; when mismatched by ~20% (e.g. the 3.33 GHz and 2.22 GHz cases), less than 4 spikes are produced in 10 ns and finally with significant deviation (e.g. 5 GHz and 1.82 GHz) no output spike is produced. This suggests that further investigation into the skyrmion magnetization dynamics may reveal an appropriate input amplitude (and other conditions) where the number of output spikes over a given time window can provide an estimate of the degree of synchronization.

## V. Conclusion

In this work, we studied novel nonlinear resonant dynamics of the core of a fixed skyrmion and showed that it has potential to lead to an energy efficient hybrid voltage controlled nanomagnetic device – CMOS device based circuit that can implement a "resonate and fire" neuron. The energy dissipation of such a device per spiking event can potentially be ~femto-Joules, which is 3 orders of magnitude (1000 times) less than an all CMOS implementation [58]. It can scale to much higher densities (~100 times less area) than an all CMOS implementation [58], while being able to exhibit resonance frequencies in the range of a few Giga-Hertz

Furthermore, future work on combining such voltage controlled nanomagnetic frequency and phase synchronization detectors with voltage controlled nanomagnetic oscillators (not discussed in detail this paper) can lead to all voltage controlled-nanomagnetic devices (with some CMOS devices) based neuromorphic circuits that are potentially very energy efficient, dense and fast.

**Acknowledgement:** M.A.A., D.B. and J.A. are supported in part by the National Science Foundation CAREER grant CCF-1253370, Virginia Microelectronics Seed Grant, Virginia Quest Commercialization Grant.